\documentclass[preprint,10pt,authoryear]{elsarticle}
\usepackage{graphicx, verbatim, url}

\journal{arXiv.org}

\begin{document}
\begin{frontmatter}

\title{Collaboration in computer science: \\ a network science approach. Part I}

\author{Massimo Franceschet}

\address{Department of Mathematics and Computer Science, University of Udine \\
           Via delle Scienze 206 -- 33100 Udine, Italy \\
           \texttt{massimo.franceschet@uniud.it}}

\begin{abstract}
Co-authorship in publications within a discipline uncovers interesting properties of the analysed field. We represent collaboration in academic papers of computer science in terms of differently grained networks, including those sub-networks that emerge from conference and journal co-authorship only. We take advantage of the network science paraphernalia to take a picture of computer science collaboration including all papers published in the field since 1936. We investigate typical bibliometric properties like scientific productivity of authors and collaboration level in papers, as well as large-scale network properties like reachability and average separation distance among scholars, distribution of the number of scholar collaborators, network resilience and dependence on star collaborators, network clustering, and network assortativity by number of collaborators.

\begin{comment}
We find that the collaboration network for computer science is a widely connected small world. The node degree distribution is right-skewed and concentrated, but the connectivity of the network does not crucially depend on star collaborators. Furthermore, the network is highly transitive and assortative mixed by node degree, in accordance with most other social networks.  The conference collaboration network is more widely and densely connected than the journal counterpart, its degree distribution has a longer tail, and it is less dependent on star collaborators. On the other hand, journal collaboration establishes a stronger relationship among authors and the journal network has a more evident core-periphery structure.
\end{comment}
\end{abstract}


\end{frontmatter}

\section{Introduction}

Collaboration is a fundamental and increasingly common feature in scientific research. Collaborative research has been associated with higher productivity, with higher impact, and, ultimately, with higher quality: from an economic perspective, collaboration allows the division of labor leading to reduced costs and time saving, consent the access to scientific funding, to expensive (possibly large-scale) equipment, and to unique scientific data. From a bibliometric perspective, collaborative works are generally more visible and more cited by other scholars; moreover, they are rated higher by peer reviewers with respect to papers written in isolation, although notable exceptions exist \citep{CF10}.

\begin{comment}
Academic collaboration is fundamental in many research fields. It arises at different levels, ranging from individual scholars collaborating in writing papers to larger collaborations involving research structures or even entire world regions. Collaborating researchers can derive scientific advantages by sharing knowledge, expertise and techniques as well as jointly controlling the accuracy and the significance of results.  Furthermore, collaboration enhances the visibility of results: a multi-authored contribution is brought to the attention of a larger number of researchers through personal contacts of each contributor. Collaboration, however, has not only advantages. A collaborative work needs deep integration among co-authors, since the final result should be a coherent and uniform piece of work. If integration among authors fails, the quality of the outcome definitely declines.
\end{comment}

In this paper, we study collaboration in computer science using a network science approach. The field of \textit{network science} -- the holistic analysis of complex systems through the study of the structure of networks that wire their components -- exploded in the last decade, boosted by the availability of large databases on the topology of various real networks, mainly the Web and biological networks \citep{N10}. The network science approach has been successfully applied to analyse disparate types of networks, including technological, information, social, and biological networks.  Here, we use co-authorship in publications as a proxy for scientific collaboration and build two differently grained network representations of collaboration in computer science: an author-paper \textit{affiliation network}, which is a bipartite graph with two types of nodes for authors and papers and links running from authors to papers that they wrote. We use affiliation networks to investigate the distribution of scientific productivity and that of collaboration level. 
A coarser and highly informative alternative representation is the \textit{collaboration network}, in which the nodes represent authors and the links are collaborations between authors in publications. A collaboration network is a type of social network, since co-authorship in publication can be interpreted as a social relationship between authors: in most cases, two authors that have written a paper together do know each other quite well, at least from a scientific perspective. This is particularly true in disciplines, like computer science, where the typical paper has few co-authors and the share of single-authored papers is not large.\footnote{This is true  to a less degree for disciplines like medicine, biology and experimental physics, where the average number of authors per paper is significantly lager than in computer science. On the other hand, in arts, humanities and some social sciences, a significant share of contributions are written by a single author and hence are not collaborative works.} We study the large-scale structure of the collaboration network for computer science, investigating properties like reachability and average separation distance among scholars, distribution of the number of scholar collaborators, network resilience and dependence on star collaborators, network clustering, and network assortativity by number of collaborators. 

In the computer science publication culture, conferences are important publication sources, and journals often publish deeper versions of papers already presented at conferences.  This is a peculiarity of computer science that makes it an original research discipline: in all other sciences, indeed, journals are the primary publication source, while monographs are the standard publication type in most social sciences, arts and humanities. This singularity motivated us to analyse separately the large-scale structure of two sub-networks of the whole collaboration graph, namely the \textit{conference} and the \textit{journal} collaboration networks. It is worth observing that the role of conferences in computer science is currently heartily discussed in the computer science literature (see \citet{F10-CACM} and references therein). 

This is the first part of our investigation of collaboration in computer science using a network science approach. In the second part of our contribution, we make a longitudinal (time-resolved) study of the network properties analysed in this paper, to get a \textit{dynamic picture} of how bibliometric and collaboration patterns evolved over time in the last half-century of computer science \citep{F11b}.

\section{Related literature}

Academic collaboration has been extensively studied in \textit{bibliometrics}, the branch of information and library science that quantitatively investigates the process of publication of research achievements \citep{S83,LPS92,KM97,BG00,SO07,CF10}. Bibliometricians observed that collaboration intensity neatly varies across disciplines. The intensity of research collaboration is negligible in arts and humanities, while social scientists often work in team, but collaborations are smaller in scale and formality compared to science disciplines. By contrast, collaborative work is heavily exploited in science, in particular in physics, medicine, and biology. Collaboration is, however, moderate in mathematics, computer science, and engineering. Moreover, collaboration generally pays in terms of impact, measured with the popular bibliometric practice that tallies the number of citations that a work receives from other papers. Furthermore, collaborative works are generally valued higher by peer experts. Both impact and quality of papers are further enhanced when the affiliations of authors are heterogeneous. Interestingly, in computer science a little collaboration, but not more than that, seems fruitful to obtain more valuable papers.

Collaboration has been also investigated under the network analysis umbrella. Sociologists have the longest tradition of quantitative study of social networks \citep{M34,DGG41,WF94,S00}. However, the notion of academic collaboration network, a particular type of social network, first appeared in 1969 in a brief note by mathematician \citet{G69}. Goffman defined the \textit{Erd\H{o}s number} for a given mathematician as the length of the shortest path on the mathematics collaboration network connecting the mathematician with Paul Erd\H{o}s.\footnote{Paul Erd\H{o}s was a notably eccentric Hungarian mathematician that is currently the most prolific and the most collaborative among mathematicians. He wrote more than 1400 papers cooperating with more than 500 co-authors \citep{G97}. Erd\H{o}s was an itinerant mathematician, living most of his life out of a suitcase visiting those colleagues willing to give him hospitality in exchange for collaboration in the writing of papers (\textit{``Another roof, another proof''}, he was used to say).} The idea of Erd\H{o}s number and hence of collaboration network, however, was informally already present in the mathematics community before 1969, since in Goffman's note we can read:

\begin{quote}
\textit{I was told several years ago that my Erd\H{o}s number was 7. It has recently been lowered to 3. Last year I saw Erd\H{o}s in London and was surprised to learn that he did not know that the function v(Erd\H{o}s; .) was being considered. When I told him the good news that my Erd\H{o}s number had just been lowered, he expressed regret that he had to leave London the same day. Otherwise, an ultimate lowering might have been accomplished.
}
\end{quote}

The note of Goffman is followed by a brief series of (occasionally sarcastic) papers of colleagues of him, including one written by Paul Erd\H{o}s himself, speculating on some theoretical properties of the collaboration graph in mathematics \citep{H71,E72,O79}.

Newman was the first to experimentally study large-scale collaboration networks with the aid of modern network analysis toolkit. He analysed the structural properties of collaboration networks for biomedicine, physics \citep{N01a,N01b}, and mathematics \citep{N04}, as well as the temporal evolution of collaboration networks in physics and biomedicine \citep{N01c}. \citet{BJNRSV02} studied the evolution in time of collaboration networks in neuroscience and mathematics. The temporal dynamics of mathematics collaboration networks is also investigated by \citet{G02}. \citet{M04} studied the structure and the temporal evolution of a social science collaboration network.

As for studies concerning the computer science collaboration network, \citet{HZLG08} considered publications from 1980 to 2005 extracted from the CiteSeer digital library. The dataset consists of 451,305 papers authored by 283,174 distinct researches. The authors studied properties at both the network level and the community level and how they evolve in time. \citet{BBNDFS09} focused on the structure and dynamics of collaboration in research communities within computer science. They isolated 14 computing areas, selected the top tier conferences for each area, and extracted publication data for the chosen conferences from DBLP 2008. The dataset contains 83,587 papers, 76,598 authors, and 194,243 collaboration pairs. They used network analysis metrics to find differences in the research styles of the areas and how these areas interrelate in terms of author overlap and migration. \citet{MZLA09} made a geographical analysis of collaboration patterns using network analysis. They considered publications from 1954 to 2007 for members of 30 research institutions (8 from Brasil, 16 from North America, and 6 from Europe) and focused on the differences in collaboration habits among these geographical areas. The dataset, extracted from DBLP, contains 352,766 papers and 176,537 authors. \citet{N01a} studied the collaboration graph for computer science as well, using the NCSTRL library, a database of preprints published in computer science during 1991-2001 and submitted by 160 participating institutions. Unfortunately, as acknowledged by Newman himself, the coverage of the used dataset (13,169 papers and 11,994 authors) is rather limited  and hence the sample is not representative of the set of computing publications. Finally, the temporal evolution of the collaboration graph for the database community is studied by \citet{EL05}. The dataset is extracted from DBLP and contains 38,773 publications written by 32,689 authors from 1968 to 2003 covering 19 journals and 81 conferences closely related to the database community. Table~\ref{networks} contains a summary of network statistics for different disciplines including the results found in this paper.

Our investigation differs from the mentioned previous studies on computer science for the following reasons: 

\begin{itemize}

\item we build the largest computer science affiliation and collaboration networks ever investigated; 

\item with the support of the affiliation network representation of collaboration, we study bibliometric properties for computer science, like author productivity and collaboration level in papers;

\item with the aid of the collaboration network we study meaningful large-scale network properties; in particular, the size of biconnected components, the concentration of collaboration using the Lorenz curve and the Gini coefficient, the collaboration network resilience and dependence on star collaborators have never been examined before for computer science;

\item we investigate separately the networks emerging from scholar collaborations in conference and journal papers.

\end{itemize}

\begin{table}
\begin{center}
\begin{footnotesize}
\begin{tabular*}{1\textwidth}{@{\extracolsep{\fill}}llrrrrrrrrr}
\textbf{disc} &  \textbf{source} & \textbf{nodes} & \textbf{edges} & \textbf{deg} & \textbf{com} & \textbf{dis} & \textbf{dia} &  \textbf{tra} & \textbf{clu} & \textbf{mix}   \\ \hline
MAT & Mat.\ Rev.\ & 253,339 & 496,489 & 3.92 & 0.82 & 7.57 & 27 &   0.15 & 0.34 & 0.12 \\ \hline
PHY &  arXiv & 52,090 & 245,300 & 9.27 & 0.84 & 6.19 & 20 &   0.45 & 0.56 & 0.36 \\ \hline
BIO & Medline & 1,520,251 & 11,803,064 & 15.53  & 0.92 & 4.92 & 24  & 0.09 & 0.60  & 0.13\\ \hline
NEU  & -- & 209,293 & -- & 11.54 & 0.91 &  6.00 & -- & -- & 0.76  & -- \\ \hline
SOC  & Soc.\ Abs.\ & 128,151 & -- & -- & 0.53 &  9.81 & -- & 0.19 & --  & --   \\ \hline
CS  & CiteSeer  & 283,174 & -- & 5.56 & 0.66 &  7.10 & 26 & -- & 0.63  & 0.28 \\ \hline
CS    & DBLP  & 688,642 & 2,283,764 & 6.63 & 0.85 & 6.41 & 23 & 0.24 & 0.75 & 0.17\\ \hline
CS-C  & DBLP  & 503,595 & 1,584,108 & 6.29 & 0.85 & 6.54 & 23 & 0.24 & 0.75 & 0.16\\ \hline
CS-J  & DBLP  & 356,822 &   987,059 & 5.53 & 0.77 & 7.26 & 25 & 0.37 & 0.77 & 0.30\\ \hline
\end{tabular*}
\end{footnotesize}
\end{center}
\caption{Structural properties of discipline collaboration networks. Column names are abbreviated as follows: disc (discipline: MAT (mathematics \citep{G02,N04,N10}), PHY (physics \citep{N01a,N10}), BIO (biomedicine \citep{N01a,N10}), NEU (neuroscience \citep{BJNRSV02}), SOC (social science \citep{M04}), CS (computer science \citep{HZLG08} and this paper), CS-C (computer science conferences; this paper), CS-J (computer science journals; this paper)), source (bibliographic source), nodes (number of nodes), edges (number of edges), deg (average node degree), com (percentage of the largest connected component), dis (average geodesic distance), dia (largest geodesic distance), tra (transitivity coefficient), clu (clustering coefficient), mix (assortative mixing). A dash sign indicates that the data is not available.}
\label{networks}
\end{table}

\section{Methodology}

Data were collected from \textit{The DBLP Computer Science Bibliography} (DBLP, for short) \citep{DBLP}. The DBLP literature reference database was developed within the last 15 years by Dr.\ Michael Ley at Trier University, Germany. DBLP is internationally respected by informatics researchers for the accuracy of its data. As of today, DBLP contains more than 1.6 million entries covering computer and information science. 

Each publication record in DBLP has a key that uniquely identifies the publication and a property that represents the publication type, such as journal article, conference article, book, book chapter, and thesis. Moreover, it contains a semi-structured list of bibliographic attributes describing the publication, like authors, title, and year of publication. This list veries according to the publication type \citep{Ley09}.

DBLP is particularly careful with respect to the quality of its data, and is especially sensible to the \textit{name problem}, which includes the cases of a scholar with several names (synonyms) and that of several scholars with the same name (homonyms) \citep{RWLWK06}. DBLP uses full names and avoids initials as much as possible. This reduces, but does not eliminate, the name problem. Furthermore, it uses effective heuristics on the collaboration graph to identify possible cases of synonyms or homonyms. For instance, if two lexicographically similar names are assigned to authors that have a distance of two in the collaboration graph, that is, these authors never directly collaborated in a paper but they have a common collaborator, then these names are identified as possible synonyms and they are further manually investigated. Furthermore, if the list of co-authors of an author splits in two or more clusters of highly interconnected authors, but with no collaborations among authors of different clusters, then we might have a case of homonym, and an additional manual check is performed. 

DBLP can be used free of charge. Data can be accessed using a Web interface or through automatic HTTP requests, and the entire dataset can be downloaded in XML format to run experiments on top of it.   

We downloaded the XML version of DBLP bibliographic dataset in early 2010 (637.9 MB) and filtered all publications from 1936 to 2008 inclusive.\footnote{We excluded year 2009 since for it the bibliography has not reached the same level of completeness as for previous years.} On top of this database we built the following networks:

\begin{comment}
The obtained XML database contains 1,238,458 publications, of which 740,465 (60\%) are conference papers and 476,061 (38\%) are journal papers. The remaining 2\% consists of proceedings (11992), book chapters (8348), books (1480), theses (98), and Web publications\footnote{We excluded author Web pages, which are catalogued as special WWW publications.} (14). 
\end{comment}

\begin{figure}[t]
\begin{center}
\includegraphics[scale=0.25, angle=0]{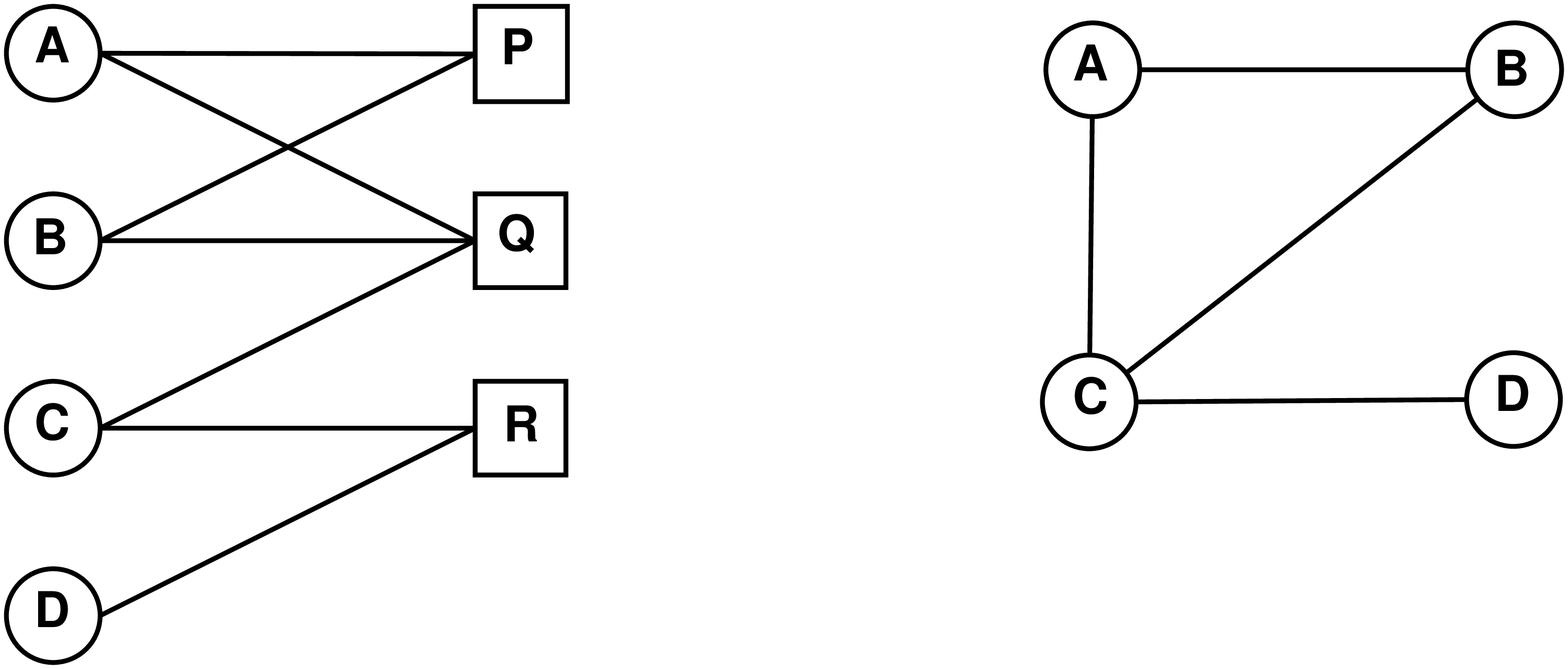}
\caption{A toy example of an author-paper affiliation network (left graph). Authors (circle nodes) match the papers (square nodes) that they wrote.  We also show the corresponding collaboration network (right graph). In this case, two authors are connected if they wrote at least one paper together.}
\label{affiliation}
\end{center}
\end{figure}

\begin{itemize}

\item \textit{Author-paper affiliation network}. This is a bipartite graph with two types of nodes: authors and papers. There is an edge from an author to a paper if the author has written the paper. See an example in Figure~\ref{affiliation}. Affiliation networks are the most complete representations for the study of collaboration \citep{N10}; in particular, on top of such bipartite representations, one can investigate both author-oriented and paper-oriented properties. The resulting affiliation network contains 731,333 author nodes, 1,216,526 paper nodes, and 3,112,192 crossing edges.

\item \textit{Collaboration network}. A collaboration network is an undirected graph obtained from the projection of the author-paper affiliation network on the author set of nodes. Nodes of the collaboration network represent authors and there is an edge between two authors if they have collaborated in at least one paper.  An example is given in Figure~\ref{affiliation}. Clearly, the collaboration network is a coarser representation with respect to the affiliation network; for instance, if three authors are mutually linked in the collaboration network, then it is not clear, from the analysis of the collaboration network alone, whether they have collaborated in a single paper or in three different ones. Nevertheless, the collaboration network is highly informative since many collaboration patterns can be captured by analysing this form of representation. Furthermore, the collaboration network is the main (mostly unique) representation of collaboration that has been studied in the network science literature. 
The resulting collaboration network contains 688,642 nodes (authors) and 2,283,764 edges (collaborations).\footnote{We excluded from collaboration networks isolated nodes, that are authors that have never collaborated. The share of these authors is about 6\% of the total number of authors.} This is, to our knowledge, the largest computer science collaboration network and the second largest discipline collaboration network ever studied, second only to the Medline collaboration network for biomedicine investigated in \citep{N01a};

\item \textit{Conference collaboration network}. In the conference collaboration network, two scholars are linked if they have collaborated in at least one conference paper. The resulting network has 503,595 nodes and 1,584,108 edges;

\item \textit{Journal collaboration network}. In the journal collaboration network, two scholars are connected if they have co-authored at least one journal paper. The resulting network contains 356,822 nodes and 987,059 edges.

\end{itemize}

We saved the collaboration networks in GraphML format (an XML syntax for graphs). We loaded them in the R environment for statistical computing \citep{R} and analysed the structure of the networks using the  R package \textit{igraph} developed by G\'{a}bor Cs\'{a}rdi and Tam\'{a}s Nepusz. On the other hand, we never materialized the (much larger) affiliation network. Instead, we used XQuery, the standard XML query language, and BaseX \citep{basex}, a light-speed native XML database, to extract the relevant properties from the XML version of the DBLP database.

\section{Analysis}

In this section we show how a structural analysis of the affiliation and collaboration networks for computer science uncovers interesting properties of the publication process of the discipline.

\subsection{Scientific productivity and collaboration level}

In this section we investigate two typical bibliometric distributions for the set of papers of a discipline: the distribution of the number of papers per author (scientific productivity) and the distribution of the number of authors per paper (collaboration level). It is worth noticing that these distributions can not be extracted from the collaboration network, since this network does not represent the links between authors and papers, but it contains only the links among collaborating authors. The correct network to investigate in this case is the author-paper affiliation network. We recall that this network is a bipartite graph with two node types representing authors and papers; edges match authors with papers they wrote. 

The distribution of the number of papers per author corresponds to the distribution of the node degree for nodes of type author in the author-paper affiliation network.  Indeed, the degree (number of adjacent nodes) of a node of type author on the affiliation network is precisely the number of papers published by the author. In substantial agreement with one of the oldest bibliometric laws -- Lotka's law of scientific productivity \citep{L26} -- the distribution of the number of papers per author is highly skewed, with most of the authors that produced a small number of contributions and few prolific ones that published a large volume of papers. In Table~\ref{Lotka} we show the relative frequency of authors that wrote a given number of papers. The table shows only the first 10 numbers of papers, but the distribution has a long tail ending at 528, the number of papers of the most prolific author. This asymmetry in scientific productivity is not characteristic of computer science but it has been noticed in many fields; for instance, \citet{M04} found a similar pattern in the productivity of social scientists, with 65.8\% of them with 1 paper, 15.1\% with 2 papers, 6.5\% with 3 papers, 3.7\% with 4 papers, 2.2\% with 5 papers, and the remaining 6.7\%  with 6 or more contributions. 

\begin{table}
\begin{center}
\begin{tabular*}{1\textwidth}{@{\extracolsep{\fill}}l|llllllllll}
\hline
\textbf{\# of papers}  &  1 &   2  &  3 &   4 &   5 &   6  &  7  &  8 &   9 &  10   \\ \hline
\textbf{\% of authors} & 53.2\% & 15.8\% & 7.7\% & 4.7\% & 3.2\% & 2.3\% & 1.8\% & 1.4\% & 1.1\% & 0.9\%  \\ \hline
\end{tabular*}
\end{center}
\caption{The scientific productivity of computer scientists. The table shows the relative frequency of authors (second row) that wrote a given number of papers (first row, from 1 to 10 papers).}
\label{Lotka}
\end{table}

As for the distribution of the number of authors per paper, it corresponds to the distribution of the node degree of nodes of type paper in the author-paper affiliation network. We found that the average computer science paper has 2.56 authors. This figure is significantly lower than the average collaboration level in other scientific fields like physics, chemistry, biology and medicine, but it is higher than the average collaboration level in social sciences and humanities (see \citet{CF10} for the collaboration level of different disciplines). Hence, computer science stays in a peculiar intermediate position where a little collaboration (2 or at most 3 scholars), but not more than that, seems to be optimal.  

Table~\ref{tab.collaboration} shows the relative frequency of papers having a given number of authors, distinguishing between conference and journal papers.  We observe that conference papers are more collaborative (2.69 authors on average) than journal papers (2.35 authors on average). In particular, notice that 19\% of the conference papers are single-author works, while this share is significantly higher, 30\%, for the papers in journals. 

\begin{table}
\begin{center}
\begin{tabular*}{1\textwidth}{@{\extracolsep{\fill}}l|llllllllll}
\hline
\textbf{\# of authors}  &  1 &   2  &  3 &   4 &   5 &   6  &  7  &  8 &   9 &  10   \\ \hline
\textbf{\% of papers} & 23.3\%  & 32.8\% &  23.5\% &  11.6\% &   4.6\% &   1.9\% &   0.8\% &   0.4\% &   0.2\% &   0.1\% 
  \\ \hline
\textbf{\% of conf.} &  19.2\%  & 32.4\%  & 25.4\%  & 13.3\%  &  5.4\%  &  2.2\%  &  0.9\%  &  0.4\%  &  0.2\%  &  0.1\%    
  \\ \hline
\textbf{\% of jour.}  & 29.7\%  & 33.4\%  & 20.5\%  &  9.0\%  &  3.4\%  &  1.4\%  &  0.7\%  &  0.4\%  &  0.2\%  &  0.1\%  
  \\ \hline
\end{tabular*}
\end{center}
\caption{The collaboration level of computer science papers. The table shows the relative frequency of papers (second row for all papers, third row for conference papers, and fourth row for journal papers) having a given number of authors (first row, from 1 to 10 authors).}
\label{tab.collaboration}
\end{table}

\subsection{Connected components} \label{components}

A \textit{connected component} of an undirected graph is a maximal subset of nodes such that any node in the set is reachable from any other node in the set by traversing a path of intermediate nodes. A connected component of a collaboration graph is hence a maximal set of authors that are mutually reachable through chains of collaborators.

It is reasonable to assume that scientific information flows through paths of the collaboration networks; we expect, indeed, that two authors that collaborated in some paper are willing to exchange scientific information with a higher probability than two scholars that never collaborated. Having a large connected component in the collaboration graph, of the order of the number of scholars, is a desirable property for a discipline that signals its maturity: theories and experimental results can reach, via collaboration chains, the great majority of the scholars working in the field, and thus scholars are scientifically well-informed and can incrementally build new theories and discover new results on top of established knowledge. Furthermore, the connectedness of a discipline is welcome in the view of proofs of theorems (and validation of experimental results) as a social human process and a community project \citep{DLP79}. Of course, collaboration represents only one way to spread scientific information; the processes of journal publishing and conference attendance make also notable contributions in this direction.

On the other hand, a high level of discipline connectedness might also have negative effects, since it favors the globalization and the standardization of results, and hence the publication of mainstream contributions at the expense of more innovative papers that explore research directions outside the established core subjects. Moreover, the independent discovery of the same theories and results by different groups of scientists, which is more likely when the discipline community is disconnected, increases the confidence of the whole community in the validity of these theories and results.

The computer science collaboration network is widely connected. The largest component counts 583,264 scholars, that is 85\% of the entire network. It is a \textit{giant component}, since it collects the great majority of nodes. There are two second largest components, the size of which, only 40 nodes, is negligible compared to that of the giant component. The third largest component has 30 nodes, and there are components for each size smaller than 30. In total we have 34,691 connected components, most of which have small sizes: 18,244 of them have size 2, while 8354 have size 3, and 3854 have size 4; hence 88\% of the components have size at most 4. The distribution of the size of the connected components that are different from the giant component has a long tail in which most components have small size and a few of them have large size.

Interestingly, the relative size of the giant component of the collaboration network for computer science matches quite well that for physics, and it is a bit higher than that for mathematics. With respect to computer science, the networks for biomedicine and neuroscience are better connected, while the cohesion of social science is lower (see Table~\ref{networks}). This means that research collaboration is more effective in medical disciplines than in social as well as hard sciences. 

\begin{comment}
The figure found by \citet{HZLG08} for computer science using the CiteSeer dataset is significantly smaller that our finding and rather odd when compared with other similar disciplines. This might indicate that the coverage of the CiteSeer dataset is less representative of the publication dataset of the computing community.
\end{comment}

A \textit{biconnected component} of an undirected graph is a maximal subset of nodes such that for each pair of nodes there are \textit{two} independent (disjoint) paths connecting them. It follows that the removal of a single node from a biconnected component does not destroy the connectivity of the component. A biconnected component is hence more tightly connected than a connected component. Information flowing on a biconnected component has more chance to reach a target node of the component since there exist two independent paths from any component node to the target node.

The largest biconnected component of the collaboration graph for computer science counts 418,001 nodes, or 61\% of the entire network, and it covers a share of 72\% of the largest connected component. The second-largest biconnected component has only 32 nodes. As a comparison, the physics collaboration has a biconnected component of 59\% of the entire network \citep{NG08}, and for social science the biconnected component occupies a share of 23\% of the network space.

\begin{comment}
\begin{figure}[t]
\begin{center}
\includegraphics[scale=0.40, angle=-90]{cc.eps}
\caption{The distribution of the size of the non-giant connected components of the collaboration graph. The size of the component is given on the $x$ axis (from 2 to 40), and the number of components of that size is plotted on the $y$ axis.}
\label{cc}
\end{center}
\end{figure}
\end{comment}

The conference collaboration network is also well connected, with 429,193, that is 85\% of the authors belonging to the giant component. The distribution of the size of secondary components has a long tail, with the second largest component counting 44 nodes. The journal collaboration network is somewhat less connected: 273,861, that is 77\% of the authors lie in the giant component. Again, smaller components distribute with a long tail in terms of size, with a second largest component of 37 elements. Hence, the conference collaboration network is more connected than the journal counterpart, indicating that information has a broader reach when flowing via conference collaboration links.

\subsection{Geodesic distances} \label{geodesic}

A high level of connectedness in the collaboration network means that scientific information -- theorems and experimental results -- can reach almost the whole community via collaboration paths. Connectedness, however, does not tell us the whole story, since it says nothing about how fast the information flows. Information flows faster along shorter paths.  In this respect, there exists a substantial difference if the average path connecting two scholars has length, say, six edges, or one hundred links.

We may assume that information preferentially flows along \textit{geodesics}, which are shortest paths in terms of number of edges on a graph.\footnote{The terms geodesic comes from geodesy, the science of measuring the size and shape of Earth; in geodesy a geodesic is the shortest route between two points on the Earth's surface.} A \textit{geodesic distance} between two nodes is defined as the length (number of edges) of any geodesic (shortest path) connecting the nodes  -- notice that a geodesic is not necessarily unique. The average geodesic distance is the mean geodesic distance among all pairs of nodes of a graph. If the graph is not connected, then there are node pairs that are not reachable. In this case the mean is typically computed on the subset of connected pairs only. The largest geodesic distance in the graph is called the \textit{diameter} of the graph. It tells us how far are two connected nodes in the worst case.

We computed the geodesic distances for all pairs of nodes in the computer science collaboration network and took the average over the subset of connected pairs (the pairs with a defined distance). Since the graph has 688,642 nodes, the number of node pairs is 237,113,557,761, of which 72\% are connected by a path.\footnote{The computation of all-pairs shortest paths is computational intensive. In the unweighted case, it takes $O(n m)$, where $n$ and $m$ are the number of nodes and the number of edges of the graph, respectively. Notice that our collaboration graph is sparse, being $m \simeq \, 3.32 n$, hence the computational complexity is of the order of $n^2$. Using the igraph R package, the computation took more than 65 hours, that is 1 microsecond per pair of nodes on average.} This figure matches the relative size of the giant component, which was found to be about 0.85 (see Section~\ref{components}). Indeed, if we randomly pick two nodes in the graph, the probability that they fall in the giant component is $0.85^2 \simeq 0.72$. Since the sizes of the other components are negligible compared to that of the giant component, this probability is a close approximation of the probability that two nodes are connected by a path.

\begin{figure}[t]
\begin{center}
\includegraphics[scale=0.50, angle=-90]{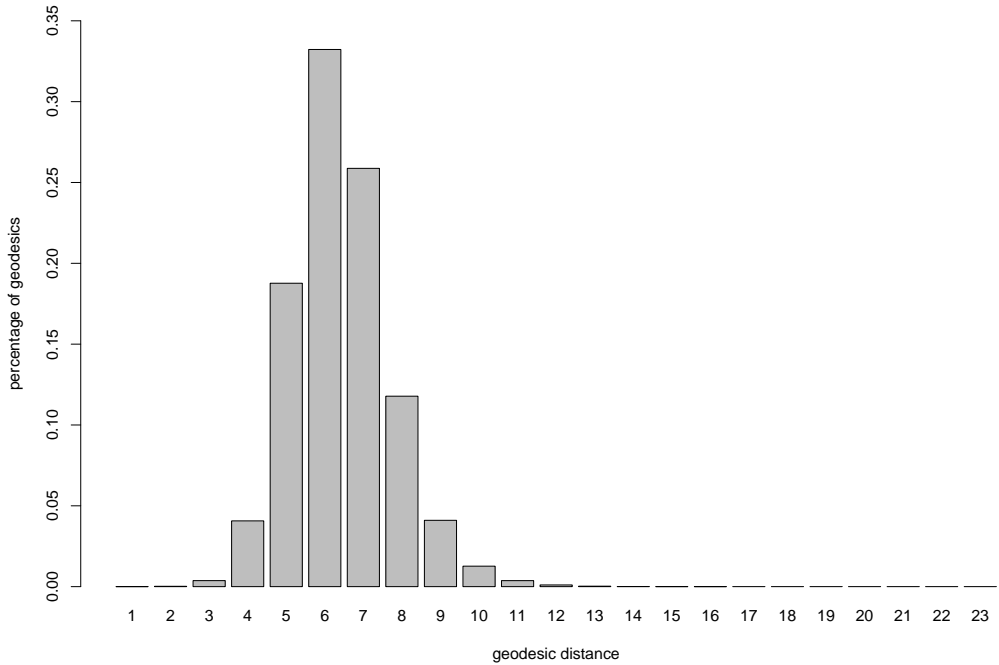}
\caption{The distribution of geodesic distances. The distances are shown on the $x$ axis (from 1 to the network diameter 23) and the height of the corresponding bar is the percentage of geodesics with that distance.}
\label{distances}
\end{center}
\end{figure}

Figure~\ref{distances} shows the share of geodesics having a given length. Notice that geodesics have typically very short lengths compared to the number of nodes: 19\% of geodesics have length 5, 33\%  have length 6, and 26\% have length 7. The average geodesic distance is 6.41, and, interestingly, distances normally distribute around this peak. The largest distance, the diameter of the computer science collaboration graph, is also remarkably small: 23 (there are 8 different geodesics with this length). Hence, computer scientists are separated on average by 6 collaboration links, a figure that matches well the legendary 6 degrees of separation found by the experimental psychologist Stanley Milgram in the 1960s with his popular small-world experiment \citep{Mi67}. These are additional good news for the computing community: not only the collaboration network is mostly connected, but the average distance is  short, and the longest one is not that longer. This means that scientific information can spread \textit{quickly} on the great majority of the computing community through its collaboration network.

The fact that distances normally distribute is  interesting because it means that the average distance of 6 links represents a typical value of all distances in the network. Furthermore, since the distribution of distances drops off rapidly around the mean, the time-consuming computation of the exact average distance can be approximated by computing the average distance on a relatively small random sample of node pairs. To demonstrate this, we estimated the average distance on a sample of 10,000 node pairs belonging to the giant component of the network. The outcome is  extremely close to the real distance of 6.415: the approximated distance is 6.427, with a 95\% confidence interval of [6.401, 6.453].

Is the computer science collaboration network a \textit{small world}? \citet{WS98} define a social network a small world if typical distances grow roughly logarithmically in the number of nodes of the network. More precisely, a network of $n$ nodes and $m$ edges  is a small world if the average geodesic distance is roughly $d = \log n / \log k$, where $k = 2m / n$ is the average node degree. Plugging into the formula the corresponding values for our collaboration network we have $k = 6.63$ and $d = 7.10$. Recalling that we measured an average distance of $6.41 < 7.10$, we conclude that the collaboration network of computer science is indeed a small world.

Comparing the found mean geodesic distance for computer science collaborations with that of other disciplines (see Table~\ref{networks}),  we notice that the separation distance for computer science is comparable with that for physics and neuroscience. Moreover, biomedicine has a lower collaboration distance, indicating that collaborations in this field are more densely intertwined. On the other hand, mathematics and, in particular, social science collaboration distances are higher, meaning that collaborations in these disciplines are less frequent and less effective.

The conference collaboration network matches well the whole network in terms of geodesic distances. A share of 73\% of node pairs are connected with an average geodesic distance of 6.54  and a largest geodesic distance of 23. In particular, 31\% of all shortest paths have length exactly 6. On the other hand, separation distances on the journal collaboration network are larger: the typical distance is now 7.26 and the largest is 25. The largest share of paths, 27\%, have length 7. Hence, scholars on the journal collaboration network have on average 7 degrees of separation instead of 6. Furthermore, only 59\% of the node pairs are connected. Summing up, the conference collaboration network is not only more widely but also more densely connected than the journal counterpart.

\citet{KSS97} have proposed to use paths on social networks among scientists as \textit{referral chains} to establish contacts with domain experts. In the simplest case, suppose I am searching for a piece of information and I am aware that you are a domain expert that most likely can answer my query. If I do not know you personally, it might be useful to know that we have a common collaborator that can arrange an introduction. In general, as we have seen, there exists a referral chain of intermediate collaborators connecting almost any scholars in computer science. Furthermore, this chain is short in the average case.  We might use the people in this chain in order to smoothly get in contact with the target scientist. To be sure, the chance of success depends on the path length but also on the strength of the intermediate path links. If, for instance, I can reach you through either collaborator A or collaborator B, and I have published with A a number of 100 papers and with B just one article, common sense suggests to use A as a broker. In other terms, one might reasonably argue that the intensity of the scientific relationship between two scholars is proportional to the number of papers they have written together. We can implement such an intuition by labelling each edge $(x,y)$ of the collaboration graph with a positive weight $1/k$, where $k$ is the number of papers that $x$ and $y$ have written together.\footnote{\citet{N01b} proposes to consider also the cardinality of the author set of the co-authored papers in order to define the collaboration weight. The plausible intuition is that the intensity of the scientific relationship is higher if two scholars collaborated on a paper in which they are the sole authors than if they wrote the paper with many other collaborators. We do not consider this factor, however, since the typical computer science paper has a small number of authors (typically 2), much smaller than in other experimental sciences in which tens or even hundreds of names can sign a paper.}  The edge label can be interpreted as a scientific distance among scholars: the more papers two authors have written together, the closer they are scientifically.

The weighted collaboration graph naturally induced the notion of \textit{weighted geodesic}: a shortest path in terms of path weight, defined as the sum of the weights of the path edges. The \textit{weighted geodesic distance} is hence the weight of a weighted geodesic. Notice that a weighted geodesic is not necessarily unique. Moreover, we expect that it differs from its unweighted counterpart. This opens an interesting question connected to the above mentioned referral chain issue: is preferable a short and weighty path, or a longer and lighter one in order to reach a given domain expert? Notice that a short path has the obvious advantage of having few intermediate scholars to bother, but a light path is desirable since the intermediate links are stronger and more reliable.

In order to investigate whether weighted and unweighted shortest paths are significantly different on the computer science collaboration network, we conducted the following experiment. We extracted a random sample of 10,000 node pairs belonging to the giant component of the computer science collaboration network and computed, for each pair of nodes, the weighted geodesic distance as well as the length of the weighted geodesic. The average weighted geodesic distance is 3.15, and the average length of the weighted geodesics is 11.27. Hence, light paths are much longer, almost twice longer, than the typical geodesic path, which is about 6 edges long. We furthermore generated 10,000 random node pairs belonging to the giant component and computed, for each of them, the (unweighted) geodesic distance as well as the weight of the geodesic. The average geodesic distance is 6.43, and the average weight of the geodesics is 5.07. Thus, short paths are significantly heavier than the typical weighted geodesic path, which weights about 3. As conjectured, for the computer science collaboration network, weighted shortest paths and unweighted shortest paths are different referral chains; the information seeker is hence faced with the dilemma of following short and weighty (unreliable) chains or long and light (reliable) paths in order to get in contact with the coveted expert.

\subsection{Node degree distribution} \label{degree}

A property of the full-scale structure of a network that is typically investigated is the distribution of the network node degrees. We recall that the \textit{degree} of a node is the number of neighbours of the node. In a collaboration network, the degree is the number of unique collaborators of a scholar. For any natural number $k$, the quantity $p_k$ is the fraction of nodes having degree $k$. This is also the probability that a randomly chosen node in the network has degree $k$. The quantities $p_k$ represent the \textit{degree distribution} of the network.

In most real networks, the degree distribution is highly \textit{right skewed}: most of the nodes (\textit{the trivial many}) have low degrees while a small but significant fraction of nodes (\textit{the vital few}) have an extraordinarily high degree. A highly connected node, a node with remarkably high degree, is called \textit{hub}. Since the probability of hubs, although low, is significant, the degree distribution, when plotted,  shows a \textit{long tail}, which is much fatter than the tail of a Gaussian or exponential model.

This asymmetric shape of the degree distribution has important consequences for the processes taking place on networks. The highly connected nodes, the hubs of the networks, are generally responsible for keeping the network connected. In other words, the network falls apart if the hubs are removed from the network. On the other hand, since hubs are rare, a randomly chosen node is most likely not a hub, and hence the removal of random nodes from the network has a negligible effect on the network cohesion. Substantially, networks with long tail degree distributions are resilient to random removal of nodes (failure) but vulnerable to removal of the the hub nodes (attack). In Section~\ref{resilience} we will investigate the resilience of the collaboration network under removal of nodes.

\begin{comment}
For instance, consider the \textit{configuration model}, a generalization of the random graph model in which one specifies a particular, arbitrary degree distribution and then forms a graph that has that distribution but is otherwise random. In a graph generated according to this model, a giant component exists if and only if
\begin{equation}
\frac{\langle k^2 \rangle - \langle k \rangle}{\langle k \rangle} > 1
\end{equation}
where where $k_i$ is the degree of node $i$, $\langle k \rangle = \sum_i k_{i}$ is the mean degree, and $\langle k^2 \rangle = \sum_i k_{i}^{2}$ is the mean-square degree (see \citep{N10}, page 456). If the degree distribution has a long tail, the mean-square degree $\langle k^2 \rangle$ turns out to be large compared to the mean degree $\langle k \rangle$, so that the above disequation is easily satisfied and most of the nodes of the network are connected.
\end{comment}

Hubs are also important for the spread of information or of any other quantity flowing on the network. In fact, hubs play a dual role in information diffusion over the network: on the one hand, since they are highly connected, they quickly harvest information, on the other hand, and for the same reason, they effectively spread it. In a network with hub nodes, the probability  that each node spreads the information to its neighbours need not be large for the information to reach the whole community.

\begin{comment}
Consider again the configuration model. Imagine a scenario in which nodes receive some kind of information and might decide to spread it over the network through their adjacent nodes.  Suppose each node that received the information communicates it with independent probability $r$ to its immediate neighbours. Is this information going to become widely popular over the network? This happens if and only if (see \citep{N10}, page 646)
\begin{equation}
r > \frac{\langle k \rangle}{\langle k^2 \rangle - \langle k \rangle}
\end{equation}
Again, for networks with long tail degree distribution the value $\langle k^2 \rangle$ is big compared to $\langle k \rangle$, and hence the probability $r$ that each individual spreads the information need not be large for it to reach the whole community.
\end{comment}

\begin{figure}[t]
\begin{center}
\includegraphics[scale=0.40, angle=-90]{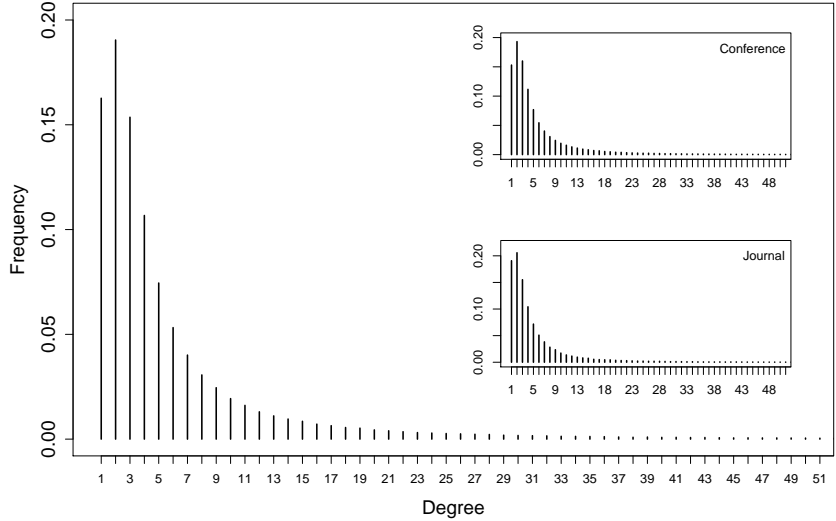}
\caption{The degree distribution for computer science collaboration network. Insets: the degree distributions for conference (top) and journal (bottom) collaboration networks.}
\label{fig:degree}
\end{center}
\end{figure}

The degree distribution for the collaboration network in computer science is depicted in Figure~\ref{fig:degree}. The distribution has in fact a long tail: roughly half of the scholars have one, two, or three unique collaborators. The other half of the scholars distribute over a slow decreasing long tail. There are, for instance, 350 scholars with 50 collaborators, and 40 scholars with 100 collaborators. The tail is in fact longer than shown in the figure, with 28 authors with more than 300 collaborators and the most collaborative computer scientist with 595 unique co-authors. The degree distributions for conference and journal articles show a similar pattern (see insets of Figure~\ref{fig:degree}), but the tail for the journal degree distribution is shorter (maximum degree is 260) than the conference counterpart (maximum degree is 481).

To quantitatively study the asymmetry of the degree distribution, we investigate the skewness and the concentration of the distribution. \textit{Skewness} measures the symmetry of a distribution. A distribution is symmetric if the values are equally distributed around its mean, it is right skewed if it contains many low values and a relatively few high values, and it is left skewed if it comprises  many high values and a relatively few low values. As a rule of thumb, when the mean is larger than the median the distribution is right skewed and when the median dominates the mean the distribution is left skewed. The mean degree for the whole collaboration network is 6.63, it is 6.29 for the conference collaboration network, and it is 5.53 in the journal case. The median degree is always 3, and the 3rd quartile is 7 for the whole and conference networks, and 6 for the journal network. A numerical indicator of skewness is the third standardized central moment of the distribution: positive values for the skewness indicator correspond to right skewness, negative values correspond to left skewness, and values close to 0 mean symmetry. The skewness indicator is 8.04 for the whole collaboration network, 7.64 for the conference network, and 6.02 for the journal network. It follows that the analysed degree distributions are right skewed, and the conference distribution is more asymmetric than the journal counterpart.

\begin{figure}[t]
\begin{center}
\includegraphics[scale=0.40, angle=-90]{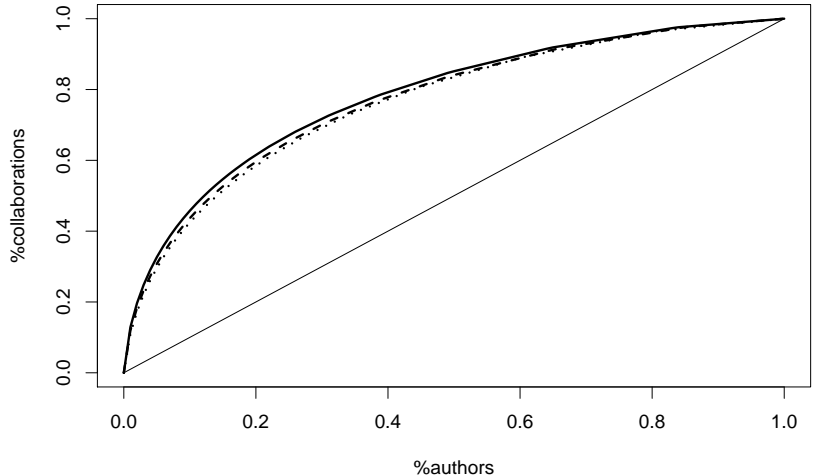}
\caption{The Lorenz collaboration concentration curves for the whole (solid curve), conference (dashed curve), and journal (dotted curve) collaboration networks. The share of most collaborative scholars  collecting a given percentage of collaboration is plotted.}
\label{fig:lorenz}
\end{center}
\end{figure}

\textit{Concentration} measures how the character (in our context, the collaborations) is equally distributed among the statistical units (the scholars). The two extreme situations are equidistribution, in which each statistical unit receives the same amount of character (each scholar has the same number of collaborators) and maximum concentration, in which the total amount of the character is attributed to a single statistical unit (there exists a super-star collaborator that co-authored with all other scholars, and each other scholar collaborated only with this super-star). We analyse the concentration of collaborations among computer scientists, that is, the concentration of edges attached to nodes in the collaboration graph. Figure~\ref{fig:lorenz} depicts the Lorenz concentration curves representing the concentration of collaboration in the whole, conference, and journal networks. Each concentration curve is obtained by sorting scholars in decreasing order with respect to the number of collaborators. Then, the share of most collaborative scholars (or network nodes) collecting a given percentage of collaboration (or network edges) is plotted. It is clear that the concentration of collaboration is far from the equidistribution situation, which is illustrated by the straight line with slope 1. For instance, the most collaborative 1\% of the scholars harvest 13\% of the collaborations, the 5\% of them collect one-third (33\%) of the collaborations, and the 10\% of them attract almost half (46\%) of the collaborations.
%This means that removing the most collaborative 10\% of the scholars from the network (the hubs of the collaboration network) implies eliminating almost half of the edges from the collaboration graph, with Draconian consequences for the connectivity of the network.
A numerical indicator of concentration is the \textit{Gini coefficient}, which is the ratio between the area contained between the concentration curve and the equidistribution line and the area representing maximum concentration. The index ranges between 0 and 1 with 0 representing equidistribution and 1 representing maximum concentration. The Gini coefficient is 0.56 for the whole network, 0.54 for the conference network, and 0.53 for the journal network. Notice that journal collaborations are slightly less concentrated than conference collaborations, but the three concentration curves are very close.

\begin{figure}[t]
\begin{center}
\includegraphics[scale=0.40, angle=-90]{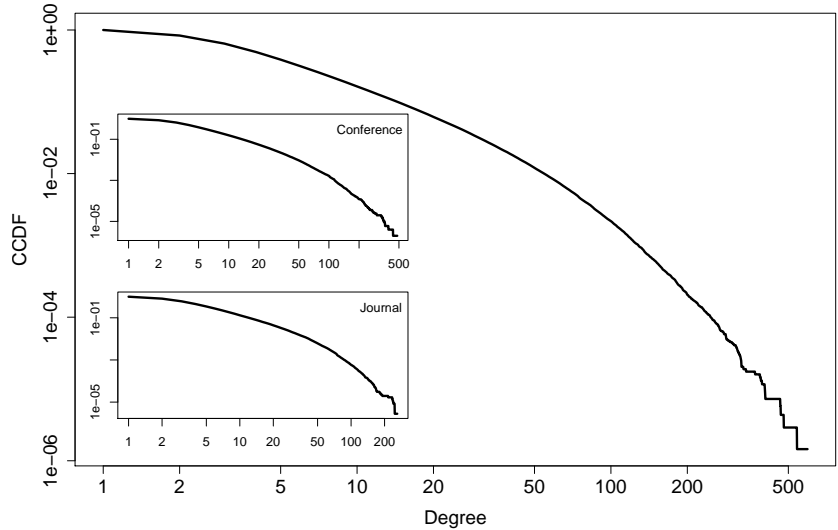}
\caption{The complementary cumulative distribution function for the degree distribution of the whole collaboration network. Insets: the same for the conference collaboration network (top) and for the journal collaboration network (bottom).}
\label{fig:cdf}
\end{center}
\end{figure}

To be sure, the most popular long tail probability distribution is the \textit{power law}. For a degree distribution, it states that the probability $p_k$ of having a node with $k$ neighbours is $C k^{-\alpha}$, where $C$ is a normalization constant and $\alpha$ is an exponent parameter. A convenient method to visualize and detect a power law behaviour is to plot the complementary cumulative distribution function (CCDF) on log-log scales (both axes are on logarithmic scales). If a distribution follows a power law, then so does the CCDF of the distribution, but with exponent one less than the original exponent (see \citep{N10}, page 252). When plotted on log-log scales, a power law appears as a straight line. Figure~\ref{fig:cdf} plots the CCDFs for the networks at hand; all show a clear upward curvature, a sign that they do not match the power law model on the entire domain.

In practice, few empirical phenomena obey power laws on the entire domain. More often the power law applies only for values greater than or equal to some minimum location. In such case, we say that the \textit{tail} of the distribution follows a power law. \citet{CSN09} developed a principled statistical framework for discerning and quantifying power law behaviour and analysed 24 real-world data sets from a range of different disciplines, each of which has been conjectured to follow a power law distribution in previous studies. Only 17 of them passed the test with a p-value of at least 0.1, and all of them show the best adherence to the model when a (limited) suffix of the distribution is considered.  We applied the techniques developed by \citet{CSN09} to detect a power law behaviour in the degree distribution of the computer science collaboration network. As expected, the degree distributions do not follow a power law on the entire regime. Nevertheless, the degree distribution for the whole collaboration network has a power law tail starting from degree 111 ($\alpha = 4.4$, p-value = 0.11). The tail contains, however, only 1098 highly collaborative scholars, which correspond to 0.16\% of all authors. The degree distribution for the conference network does not follow a power law in any significant portion of its tail. Finally, the degree distribution for the journal network matches a power law from degree 105 ($\alpha = 5.79$, p-value = 0.67); the tail is 178 scholars long, or 0.05\% of the entire distribution. There is, however, a longer (551 scholars, 0.15\%) but statistically less significant power law distributed tail starting from degree 77 ($\alpha = 4.91$, p-value = 0.09). All in all, two of the analysed networks (the whole and the journal one) have a power law distributed tail, but the relative size of the tail is in both cases rather limited. We conclude that the process of \textit{preferential attachment}  -- the attitude of scholars to collaborate preferentially with highly collaborative peers, which is one of the possible causes for the power law behaviour \citep{P76,BA99}, is not a valid explanation for the generation of the computer science collaboration network.

\subsection{Network resilience} \label{resilience}

\textit{Percolation}\footnote{The name comes from percolation studies in physics.} is one of the simplest \textit{processes} taking places on networks. The process progressively removes nodes, as long as the edges connected to these nodes, from the network, and studies how the connectivity of the network changes. In particular, one wants to find the fraction of nodes to remove from the network in order to disintegrate its giant component into small disconnected pieces. If such a fraction is relatively large, then the network is said to be resilient (or robust) to the process of percolation. Our study of percolation on collaboration networks is shaped by the following questions:

\begin{itemize}

\item What is the best removal strategy to destroy the overall connectivity of the collaboration network?

\item What is the tipping point in the percolation process after which the network consists of only small disconnected clusters?

\item Are the most collaborative scholars responsible for  keeping the network connected?

\end{itemize}

To address the above questions, we performed the following computer experiment. We implemented a computer procedure that progressively removes nodes from the collaboration network. At each step the procedure removes an increasing number of nodes from the original network and, after each removal, it computes the relative size of the giant (largest)  component of the resulting sub-network. More precisely, the procedure initially generates, according to a given strategy that will be discussed below, a number of $5000 \cdot 20 = 100,000$ nodes (about 15\% of the total number of nodes) to remove from the original graph. At each step $i$ from $1$ to $20$, the procedure removes the first $n_i = 5000 \cdot i$ nodes of the generated ones from the original network as well as the incident edges and computes the share of the giant component of the resulting network.
The procedure can choose the removal nodes according to three strategies: (i) \textit{random-driven percolation}, in which randomly chosen nodes are removed, (ii) \textit{degree-driven percolation}, in which the nodes are removed in decreasing order of node degrees, and (iii) \textit{eigenvector-driven percolation}, in which the nodes are removed in decreasing order of node eigenvector centrality scores.\footnote{Eigenvector centrality scores correspond to the values of the dominant eigenvector of the graph adjacency matrix. These scores may be interpreted as the solution of a linear system in which the centrality of a node is proportional to the centralities of the nodes connected to it.} To avoid possible biases in  random-driven percolation due to the non-deterministic output of the random generator, we repeated the experiment a large number of times and took the average of the results.\footnote{A small bias might be introduced also in the other two cases, whenever one has to partially remove nodes with the same degree or eigenvector centrality. We did not consider this issue in our experiment.}

\begin{figure}[t]
\begin{center}
\includegraphics[scale=0.40, angle=-90]{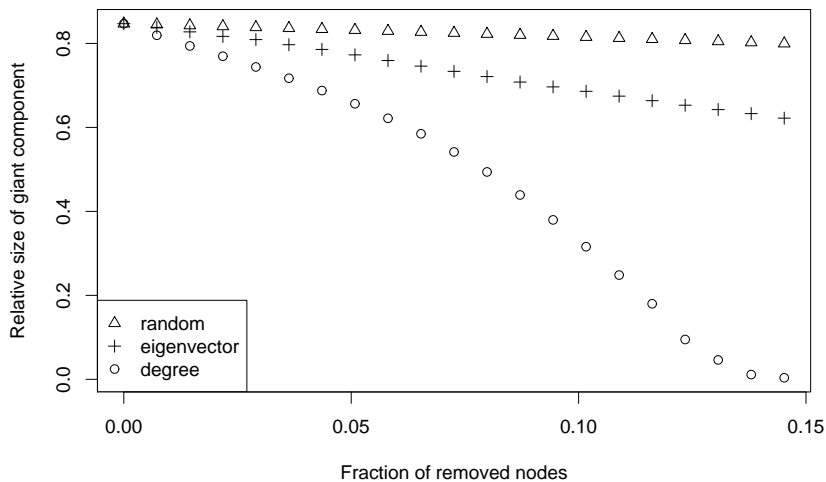}
\caption{Effect of percolation on the connectivity of the computer science collaboration network. The relative size of the giant component is shown as an increasing fraction of nodes are removed according to three different percolation strategies.}
\label{fig:robust}
\end{center}
\end{figure}

\begin{figure}[t]
\begin{center}
\includegraphics[scale=0.40, angle=-90]{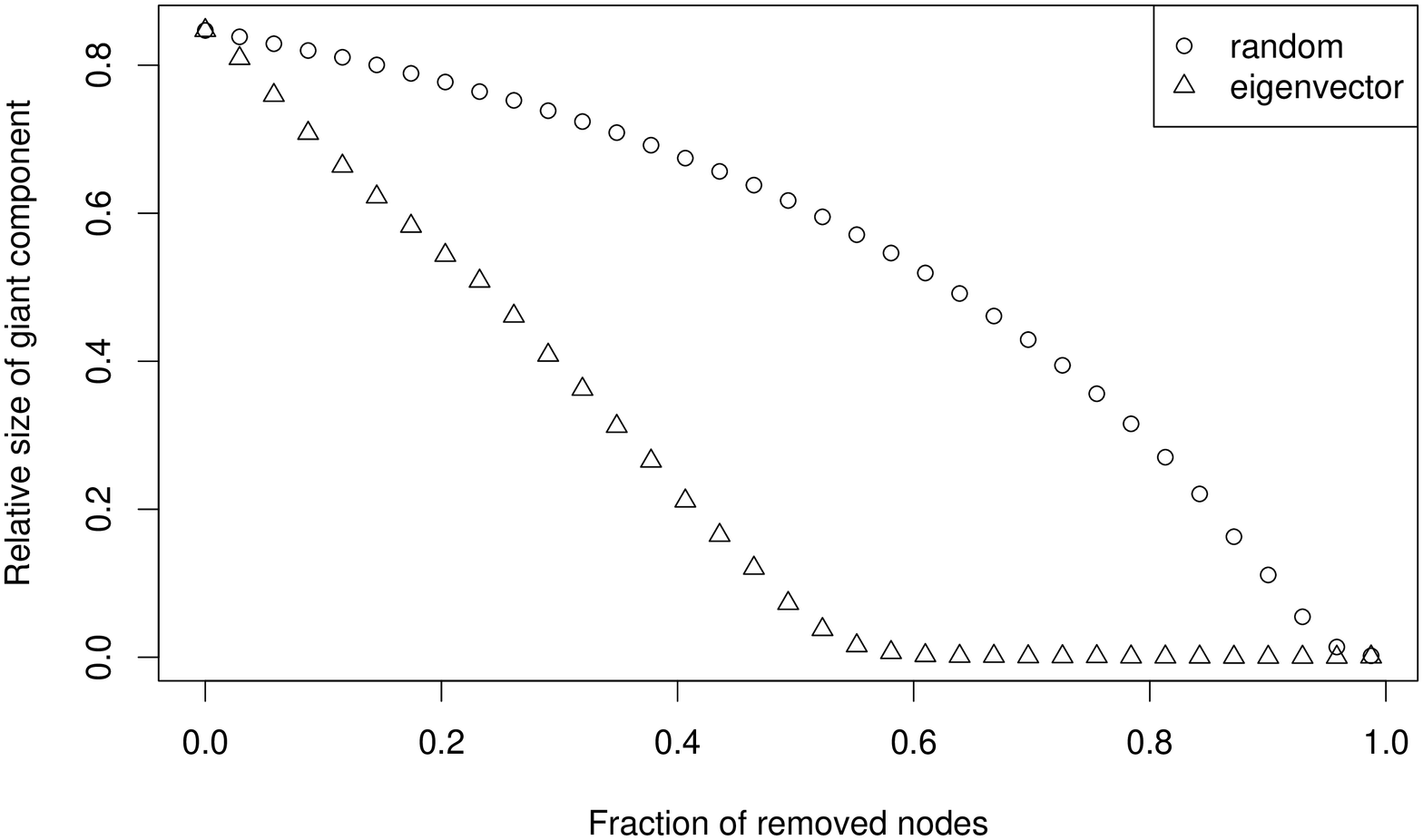}
\caption{Effect of random-driven and eigenfactor-driven percolation on the connectivity of the computer science collaboration network.}
\label{fig:robust2}
\end{center}
\end{figure}

Figure~\ref{fig:robust} clearly shows that the most effective strategy (among the surveyed ones) to destroy the connectivity of the collaboration network is to percolate the highly collaborative scholars: after the removal of about 12\% of the most collaborative scholars, the giant component of the collaboration network, which initially contains 85\% of the nodes, falls below 10\%, and it soon vanishes when the removal fraction is a little higher (15\%). The effect of removing scholars with high eigenvector centrality is much less substantial: the largest component is still considerable, about 62\% of the network, when a fraction of 15\% of the most central nodes are removed. Finally, the random removal of nodes has a negligible effect on the giant component of the network: after the percolation of a share of 15\% of randomly chosen nodes the network is still highly connected, with 80\% of the nodes belonging to the giant component. For a comparison, \citet{N10} found that to destroy the connectivity in the physics collaboration network it is necessary to remove a share between 20\% and 30\% among the highly collaborative physicists.

The shape of the degree-driven percolation curve shows clear curvatures. In particular, we can distinguish three main phases in the percolation process. An initial phase, up to a removal fraction of 7\%, in which the decrease of connectivity is limited (3-4 percentage points at each step). In this phase, although we severely attack the network by removing its most important hubs, the effect is somewhat reduced since the collaboration frame is still densely intertwined. That is, each node pair in the giant component is connected by more independent paths, and the removal of some of them do not prevent reachability.\footnote{As we have seen in Section~\ref{components}, the relative size of the largest biconnected component of the network is quite large.} In a following phase, which extends up to a removal share of 12\%, the reduction of connectivity is more notable (5-9 percentage points). In this phase, the size of the largest component is below 50\% and its collaboration frame is weaker and more vulnerable. Hence, the effect of the attack is more devastating. In the last segment, up to a removal percentage of 15\%, the relative size of the giant component is below 10\% and it goes rapidly to 0; in this phase the network consists of small disconnected clusters, none of which strongly dominates the others. Hence, for degree-driven percolation the tipping point after which only small disconnected clusters exist is around 15\%.

The overall effect of random-driven and eigenvector-driven percolations of network connectivity is shown in Figure~\ref{fig:robust2}. The shape of the random-driven percolation curve shows a clear upward curvature, meaning that the effect of random removal has a higher impact on connectivity when a significant fraction of the nodes have been already removed, as in the degree-driven case. Most of the nodes must be removed before connectivity is lost, and the tipping point where the network falls apart into small pieces is around 90\%, far beyond the percolation threshold for the degree-driven case.  The eigenvector-driven percolation curve is instead linear up to its tipping point around 50\%, which lies between the percolation thresholds of the degree-driven and random-driven processes.

A similar analysis for the second-largest component shows that its size during the percolation process is never significant (always below 1\%). This means that, when a giant component exists, relatively small pieces belonging to the periphery of the giant component separate during the percolation process, and it never happens that the giant component splits into two fragments of similar size. Only when the network is divided into small disconnected clusters, the size of the giant and sub-giant components are comparable.

We are left to the last question posed at the beginning of the present section: are the star collaborators, the hubs of the collaboration network, responsible for the connectivity of the overall network? Our answer, maybe surprisingly, in negative. The hubs are nodes with an \textit{extraordinary} number of neighbours. Let us define hubs as those nodes with a degree higher than or equal to the 99th percentile in the degree distribution, that is the 1\% of nodes with highest degree. These are the 7036 scholars of the network with at least 54 collaborators. Recall that the average scholar has between 6 and 7 collaborators, hence hub scholars have a number of collaborators 8 times higher than the average scholar. The removal of these super-star collaborators has, in fact, a negligible effect on the connectivity of the network: the size of the giant component decreases form 85\% to 81\%. On the other hand, as we have seen, to dismantle the network we need to remove at least a share of 15\% of top collaborative scholars, that is all authors with a degree higher than 11. In our assessment, scholars with 11 unique collaborators are not collaboration hubs (I have 15 unique collaborators, and I really do not feel I am a collaboration star in computer science). This conclusion matches the findings of \citet{M04} for the social science collaboration network. Hence, the computer science collaboration network is not glued together by star collaborators and, while such actors are likely very influential within their local communities, they do not control information diffusion on the \textit{whole} computer science collaboration network.

As for conference and journal collaboration networks, the results are similar but the tipping points are lower. In particular, the tipping points for the journal network are below those for the conference network. For instance, the journal network falls to pieces after 7\% of the most collaborative scholars are inhibited, while the conference network crumbles when 9\% of the most collaborative scholars are removed. This means that the journal network is more fragile and its connectivity is more dependent on star collaborators.

\subsection{Clustering and mixing}

In common parlance, clustering, known also as transitivity, measures the average probability that ``the neighbour of my neighbour is also my neighbour''. The definition of clustering can be formalized as follows. Let a \textit{connected triple} be a triple of nodes $x$, $y$ and $z$ such that $x$ is linked to both $y$ and $z$. That is, $y$ and $z$ have a common neighbour, $x$. A \textit{triangle} is a triple of nodes such that all pairs of nodes are connected by an edge. Notice that a triangle codes for three connected triples, each one centered at one vertex of the triangle. The  \textit{global clustering  coefficient} of a network can be defined as
\begin{equation} \label{clustering1}
T = 3 \frac{N_\bigtriangleup}{N_\wedge}
\end{equation}
where $N_\bigtriangleup$ is the number of triangles and $N_\wedge$ is the number of connected triples in the network. The factor 3 constrains the coefficient to lie in the range between 0 and 1. Thus, $T=1$ implies perfect transitivity, that is, a network whose components are \textit{complete} graphs,\footnote{An undirected graph is said \textit{complete} if every pair of different nodes is connected by an edge.} while $T=0$ implies no triangles, which happens, for instance, in a network whose components are \textit{tree} graphs.\footnote{A \textit{tree} is an undirected graph that is both connected and acyclic.}

Local clustering refers to a single node. For a vertex $i$, its \textit{local clustering coefficient} $C_i$ is the fraction of neighbours of $i$ that are connected, that is the number of pairs of neighbours of $i$ that are connected divided by the number of pairs of neighbours of $i$. By taking the average local clustering over all nodes of a network we have an alternative (but different) definition of global clustering coefficient \citep{WS98}:

\begin{equation} \label{clustering2}
C = \frac{1}{n} \sum_i C_i
\end{equation}

It is worth noticing that these two definitions of clustering -- $T$ and $C$ expressed respectively by Equations~\ref{clustering1} and~\ref{clustering2} -- are not equivalent and can give substantially different results for a given network. We prefer definition $T$ because it has an intuitive interpretation -- the average probability that two neighbours of a node are themselves neighbours.  To distinguish between them, in the following we will refer to $T$ as transitivity coefficient and to $C$ as clustering coefficient.

The transitivity coefficient of the computer science collaboration network is 0.24. This means that, on average, the chance that two scholars that share a common collaborator wrote a paper together is almost one-fourth. This is a rather high probability, indeed. As a comparison, the transitivity coefficient for a random network of the same size is $9.6 \cdot 10^{-6}$, and that for a network with the same degree distribution of our collaboration network but otherwise random is $1.4 \cdot 10^{-4}$; both values are orders of magnitude lower that what we computed on the real collaboration network.\footnote{The transitivity coefficient of a random graph with $n$ nodes and $m$ edges is $m / m_{*}$, where $m_* = n (n-1) / 2$ is the maximum number of edges of the graph. The transitivity coefficient of a random graph with degree sequence $k$ is $1/n \, (\langle k^2 \rangle - \langle k \rangle)^2 / \langle k \rangle^3$, where $\langle k \rangle$ is the mean degree and $\langle k^2 \rangle$ is the mean-square degree.} This large discrepancy is a clear sign of real social effect at work in the context of academic collaboration: authors with a common collaborator have several good reasons to write a paper together, for instance they are probably working on very close topics or they scientifically know each other through the common collaborator.

The transitivity coefficient for the conference and journal collaboration networks are 0.24 and 0.37, respectively. These values might indicate that journal collaboration establishes a stronger relationship among authors, so that authors having a common journal collaborator are professionally and maybe socially closer than authors sharing a common conference collaborator, and hence more inclined to collaborate themselves. In computer science publishing a paper in a journal is, generally,  harder than writing a paper for a conference \citep{F10-CACM}, and this might explain the difference in strength between conference and journal co-authorship. Comparing with other disciplines, the magnitude of transitivity in computer science is comparable with that in sociology, larger than that in mathematics and biomedicine, and lower than that in physics (Table~\ref{networks}).

The clustering coefficient for the computer science collaboration network is 0.75 (0.75 for conference collaboration, and 0.77 for journal collaboration), much larger than the transitivity coefficient. This confirms, once again, that the two clustering measures are far different. Digging deeper, we computed the local clustering coefficient for each node and noticed that  almost half of the nodes (48\%) have local clustering coefficient equal to 1, and this  explains the large value of the clustering coefficient. Recall that a node $i$ has local clustering $C_i = 1$ if its immediate neighbours form a complete graph (a clique). Notice that the neighbourhood clique can be extended by adding node $i$ itself. We noticed that most of these special cliques have small size. The most frequent pattern (37\%) is a clique of size 3, that is a triple $i$, $j$, $k$ of scholars such that $j$ and $k$ are the only collaborators of $i$ and they are themselves collaborators (not necessarily on a paper with $i$). Moreover, 26\% of these special cliques have size 4, 15\% of them have size 5, 8\% of them have size 6, and the distribution decreases slowly with a long tail ending with size 114. This large maximum value, however, is easily explained by the existence of a paper with 114 co-authors.\footnote{A paper with such large number of authors is quite unnatural in computer science. In fact, this hyper-authored paper is an article in bioinformatics, an area at the intersection of biology and computer science.}

\begin{comment}
Notice that this is not necessarily the largest clique in the network, but only a lower bound.\footnote{Finding the largest clique in a graph is an NP-complete problem, that is, most likely, there exists no efficient (polynomial) algorithm to solve the problem.} Indeed, there might be a larger clique such that the clique is not the neighbourhood of any of the nodes in the clique.
\end{comment}

Social networks differ from most other types of networks, including technological and biological networks, at least in two aspects \citep{NP03}. First, the transitivity coefficient is higher for social networks. Second, they show positive correlation between the degrees of adjacent nodes, while other networks have negative correlation. \textit{Assortative mixing}  is the tendency of nodes to connect to other nodes that are like them in some way. In particular, assortative mixing by degree is the tendency of nodes to connect to other nodes with a similar degree. In our context, we have assortative mixing by degree if scholars collaborate preferentially with other scholars with similar number of collaborators. We have disassortative mixing by degree if collaborative scholars co-author with hermits and vice versa. We have no mixing at all if none of these patterns is clearly visible.

A quantitative measure of the magnitude of mixing by degree can be computed using the \textit{Pearson correlation coefficient} applied to the degree sequences of nodes connected by an edge. The coefficient ranges between -1 and 1, where negative values indicate disassortative mixing, positive values indicate assortative mixing, and values close to 0 indicate no mixing. The coefficient is 0.17 for the whole network, 0.16 for the conference one, and 0.30 for the journal one. The values are statistically significant. Hence, collaboration networks in computer science confirm assortative mixing by degree, as other social networks: collaborative computer scientists tend to collaborate with other collaborative computer scientists, and solitary authors match preferentially with other solitary authors. Notice the higher value for collaboration in journal papers, indicating that this collaboration pattern is stronger in this case. These findings are useful to picture the structure of the collaboration network. A network that is assortative by degree has a \textit{core-periphery structure}: a dense core of high-degree nodes is surrounded by peripheral low-degree ones.

In fact, the correlation is even stronger. Pearson correlation coefficient is an appropriate measure of correlation when the data samples roughly follow a normal distribution. In this case, the mean of the samples, which is used in the computation of the coefficient, represents a characteristic scale for the network. However, we have seen in Section~\ref{degree} that our collaboration networks are \textit{scale-free}: their degree distribution is highly right-skewed and the distribution mean does not represent a typical value of the number of collaborators of a scholar. To correct for the bias introduced in the Pearson correlation coefficient by the use of asymmetric degree distributions, we can either make a logarithmic transformation of the degree sequences before using the Pearson coefficient formula, or use a non-parametric correlation method, like the Spearman one. Both methods give the same results: the correlations increase to 0.25, 0.21, and 0.36 in the whole, conference, and journal collaboration network, respectively.

\section{Conclusions}

We have analysed collaboration in computer science using a network science approach. Substantially, we have found that the scientific productivity of computer scientists is highly asymmetric, in agreement with Lotka's law of scientific productivity. The collaboration level in computer science papers is rather moderate with respect to other scientific fields, indicating that a little collaboration of two or at most three authors is optimal in computer science. However, conference papers are more collaborative than journal ones. This suggests that collaboration is more important when there are stringent deadlines for the production of a paper, like those imposed by computer science conferences. 

The computer science collaboration network is a widely connected small world, hence scientific information flows along collaboration links very quickly and it potentially reaches almost all scholars in the discipline. This signals the reached scientific maturity of the relatively young field of computer science. The distribution of collaboration among computer science scholarsis highly skewed and concentrated, with few star collaborators responsible for a relatively high share of collaborations. The collaboration network is, however, resilient to the removal of these star collaborators, meaning that the connectivity of the network does not crucially depend on them. These is good news for the computer science community, since it means that this restricted circle of influential scholars with many contacts do not control the diffusion of information on the whole discipline, although they are probably very influential within their local communities.

Finally, while the conference collaboration network is more widely and densely connected than the journal counterpart, journal collaboration establishes a stronger social relationship among authors, also because, as observed above, the typical journal paper has fewer authors than the average conference contribution. The journal network is more dependent on star collaborators, and these highly collaborative authors prefer to collaborate with other star collaborators, leading to a core-periphery structure in the journal graph. These patterns might indicate that conferences are better to widely and quickly communicate scientific results, while journals are optimal to establish stronger and longer scientific relationships with other scholars.

\bibliographystyle{elsarticle-harv}
\bibliography{bibliometrics}

\end{document}